\documentclass[usenatbib,usegraphicx]{mn2e}
\def\lsim{\raise0.3ex\hbox{$<$\kern-0.75em\raise-1.1ex\hbox{$\sim$}}}
\def\gsim{\raise0.3ex\hbox{$>$\kern-0.75em\raise-1.1ex\hbox{$\sim$}}}

\title{ Global amount of dust in the universe}
\author[M. Fukugita]
{Masataka Fukugita\\Institute for Advanced Study, Princeton NJ08540, U.S.A. and
\\Institute for Cosmic Ray Research \& Institute for the Physics and
Mathematics of the Universe, \\ ~University of Tokyo, Kashiwa 2778582, Japan}
\date{Accepted  Received }
\pagerange{\pageref{firstpage}--\pageref{lastpage}}
\pubyear{2009}

\begin{document}

\label{firstpage}

\maketitle

\begin{abstract}
  It is pointed out that the total amount of dust in the Universe that
  is produced in stellar evolution in the entire cosmic time is
  consistent with the observed amount, if we add to the dust amount
  inferred for galactic discs the amount recently
  uncovered in galactic haloes and the surrounding of galaxies in
  reddening of the quasar light passing through the vicinity of
  galaxies. The inventory concerning the dust closes.  This implies
  that dust produced from stars should survive effectively for the
  cosmic time, and that a substantial amount of dust is produced
  in the burning phase of evolved stars of intermedaite mass.

\end{abstract}


\section{Introduction}

A significant amount of gas is ejected into interstellar space during
the life of stars by stellar wind mass loss and at supernova explosions.
Gas contains heavy elements either taken from the initial gas or 
produced during the stellar evolution. 
A significant fraction of heavy elements condenses to form dust.
Although the mechanisms to produce dust have not been well understood,
the net amount of dust produced from gas with heavy elements can be
estimated with some confidence.  

Reasonable estimates are available as to how much material is
processed by stars by integrating the star formation rate as a
function of cosmic time.  Using recent compilations of the star
formation rate, we estimate that the total amount of material (fuel)
processed lies between $\Omega_{\rm fuel}=0.004$ and 0.010 (e.g.
Fukugita \& Peebles 2004, hereafter FP04; Hopkins and Beacom 2006;
Fardal et al. 2007; Nagamine et al. 2006; Ouchi et al. 2009; Bernardi
et al. 2010; Bouwens et al. 2010), assuming the Chabrier initial mass
function (Chabrier 2003).  We refer to this amount as the fuel.  These
two numbers represent the curves that pass through close to the lower
and upper parts of the star formation rate plot\footnote{We take as
  our fiducial the Chabrier initial mass function for $M<1 M_\odot$
  and the Salpeter initial mass function for $M>1 M_\odot$. Chabrier
  takes the mass function slope $-2.30$ rather than $-2.35$ of
  Salpeter at a higher mass (he revised to $-2.35$ in his later version). 
 The difference in total mass is about 5\%,
  which is smaller than the uncertainty that concerns us here. We also
  note that the initial mass function used in FP04 is close to the
  Chabrier initial mass function, but gives the total mass 10\%
  smaller than the Chabrier.}. We remark that the uncertainty seen in the
two numbers largely arises from the normalisation in low redshift
rather than apparently more ambiguous high $z$ behaviours, for
instance, whether the star formation rate declines at $z>3$ as
favoured by recent observations (Ouchi et al. 2009; Bouwens et al.
2010).  The time span is short at higher redshift and the integrated
contribution to the total amount of stars from $z\ge 2$ is roughly
$<20$\%, so that the high redshift behaviour is not important in our
argument\footnote{The normalisation at low redshift, which varies
  among the authors, is the major source of the uncertainty. Also note
  that the star formation rate increases significantly between $z=0$
  and $z=0.1$.}.

On the other hand, stars and their remnants, white dwarfs, neutron
stars and black holes, that reside in galaxies are estimated to be
\begin{equation}
\Omega_{\rm star}=0.0030\pm 0.0005
\label{eq:stardensity}
\end{equation}
(Fukugita \& Peebles 2004; hereafter FP04; 10\% upward shift is
applied to agree with the initial mass function we use here).  
Similar estimates have been made by a number of authors
(e.g., Shankar et al. 2004; Oohama et al. 2009; Bernardi et al. 2010),
and they fall in the range indicated by the two numbers.  There is a
gap between the two numbers, $\Omega_{\rm fuel}$ and $\Omega_{\rm
  star}$. This may primarily be ascribed to the material shed by stars
during the evolution, as either stellar winds or supernova explosions,
whereas there still remains a gap between the two values is left as a
problem in the future.

Taking the initial mass function of stars given by Chabrier, we
estimate that the gas fraction shed by stars during the evolution is
0.60 times the mass locked into stars and stellar remnants, as
summarised in what follows.  Therefore, the amount of baryons consumed
in star formation is $\Omega_{\rm fuel}=0.0030\times 1.60=0.0048$,
which agrees with the amount of fuels estimated from the integration
of the star formation rate at its lower value of the integral of the
star formation rate.  In the following argument we assume the total
amount of the fuel is $\Omega_{\rm fuel}=0.005$ for the consistency of
the argument.  This leads to the full consistency of the stellar
baryon and energy budget, as noted in FP04, including the stellar
light emission, supernova rates and heavy element production, while
the accounting must be improved especially as to the star formation
rate. At the upper edge of the star formation rate, the integral of
the star formation rate seems to overshoot the stars which we see
today.

In our calculation  of the dust production we use the final
mass-initial mass relation for white dwarfs (Serenelli \& Fukugita 2007;
see also Salaris et al. 2009), the mean mass of which is
0.62$M_\odot$ for main sequence stars with the 
mass $1-8M_\odot$, when averaged over the
Salpeter mass function. For neutron stars we adopt 1.35$M_\odot$ 
for the main sequence mass $8-25M_\odot$, 
and 7.5$M_\odot$ for putative stellar black holes
for the main sequence mass $25-100M_\odot$ (Heger et al. 2003).  
The remnant black hole mass is highly
uncertain, but the fuel that ends with black hole is 8\% and 
the remnant mass is only 1.4\% the fuel mass
with our adopted black hole mass:
the large uncertainty in the remnant black hole mass is not
important for our considerations.
For completeness let us quote that 11\% of eq.(\ref{eq:stardensity}) are
partitioned into white dwarfs, 8\% are substellar and
neutron stars are 2\%. 

We take this simple model as the base to estimate the amount of dust
produced in stellar evolution.  FP04 have calculated that the observed
extragalactic background light amounts to the energy density
$\Omega=5.1\pm1.5\times 10^{-6}$ where the optical and far infrared
contribute by 2:1, with a 7\% addition by neutrinos, and that the
binding energy in heavy element we observe today is altogether
$\Omega=-5.7\pm1.3 \times 10^{-6}$, including nuclei locked in or
sequestered from stars. This nearly balances the energy in the
extragalactic background light with the error arising from the two
vastly different accounting kept in mind.  It was also shown that the
energy output from stars isexpected to be  
$\Omega=5.4 \times 10^{-6}$ for the given
fuel that amounts to $\Omega=0.005$.
We note that there is no missing metal problems at least at
$z=0$, but note that 80\% of heavy elements are locked in
white dwarfs and 25\% of \lq metals' is 
sequestered in neutron stars and black holes. 

In addition, it was also shown (FP04) that the present day rate of
core collapse supernovae $0.0079^{+0.0024}_{-0.0039} (100{\rm
  yr~Mpc)}^{-3}$ obtained from the star formation rate, integrating
over the initial mass function from 8$M_\odot$ to 100$M_\odot$ agrees
with the observed rate $0.0076^{+0.0064}_{-0.0020} (100{\rm
  yr~Mpc)}^{-3}$ albeit with large errors in the observed value.
These supernovae produce iron at $\Omega_{\rm Fe}=3.6\times 10^{-6}$.  A
similar estimate of type Ia supernovae, with their normalisation
adjusted to the present rate, gives the iron abundance $2\times
10^{-6}$, which results in the total iron abundance $\Omega_{\rm
  Fe}=6\times 10^{-6}$, when added, in agreement with the estimate of
the cosmic iron abundance $\Omega_{\rm Fe}=6.3\times 10^{-6}$ for
materials which are not locked up in stellar remnants, or sequestered 
from them.  The network of
consistencies among the numbers shown here points towards the validity
of this simple framework, concerning the hydrogen fuels consumed and
the heavy element produced.  This tempts us to apply a similar
consideration to the amount of cosmic dust, which was not done in
FP04.

It is yet poorly understood where and how dust is produced, and how
long does it survive. The calculation we show in this paper, based
basically only on the final and initial mass budgets, circumvents these
poorly understood aspects and uncertainties. It gives accounting of
dust produced, which is subject to much less uncertainties. We take
the Hubble constant $H_0=70$km s$^{-1}$Mpc$^{-1}$, matter density
$\Omega_m=0.3$ in a flat universe whenever necessary.

\section{The amount of cosmic dust}

We take the traditional elemental abundance given by Grevesse and
Sauval (2000), while noting that more recent solar abundance
estimate by Asplund et al. (2005) leads to somewhat a smaller
abundance of heavy elements.  The adoption of the new abundance
changes some details of our results, but our conclusions are unaffected.  The
initial solar abundance is $Z_\odot = 0.019$ or $Z/X|_\odot=0.027$ 
using the
Grevesse and Sauval table with the present day solar abundance
$Z/X|_{\odot\rm surface}=0.023$. 
with the aid of the solar model of Bahcall et al.  (2001).
The solar abundance would
be reduced to $Z_\odot = 0.012$ if we adopt the revision
by Asplund et al., 40\% smaller than the Grevesse and
Sauval value.

The mean metallicity correlates with luminosity of
galaxies (Tremonti et al. 2004). When we integrate the metallicity 
over the luminosity function we obtain the cosmic average 
$\langle Z\rangle=0.83Z_\odot$.

We assume that refractory elements, {\rm Si}, {\rm Fe} and {\rm Mg} in
interstellar gas or those ejected into interstellar gas all {\it eventually}
condense to solids. The condensed material is not precisely
identified, but it is argued that they form effectively Mg$_{\rm
  x}$Fe$_{\rm 2-x}$SiO$_4$ (the composition of olivine, forsterite or
fayalite) with $x\approx 1$ and Mg$_{\rm
  x}$Fe$_{\rm 1-x}$SiO$_3$ (enstatite) for a lesser amount
(Weingartner \& Draine 2001).  The solar
elemental abundance of these three refractory elements is similar,
32$-$38 ppm per hydrogen, and hence are almost saturated if all {\rm Si} is
condensed to the olivine composition. Observationally, these elements are
known to be highly ($>90\%$) depleted from interstellar gas.  We assume that
these elements condense into dust, and they take oxygen by
20\% locked up in the silicate.

The other prominent component of dust is carbonaceous material,
likely including polycyclic aromatic carbon for small grains to graphite for
large grains.  The estimate for the fraction that condense to 
carbonaceous material varies. 
We take the fraction of condensation
to be $50\pm25$\%, 
consistent with the value taken by  Weingartner \& Draine (2001) 
(Weingartner \& Draine 2001).

Adding the two components we obtain the dust to metallicity mass ratio
\begin{equation}
\eta={\rm dust}/Z\simeq 0.28\pm0.07 ,
\label{eq:dustZ}
\end{equation}
where the error dominantly arises from the two choices of the silicate
and partly from the assumed amount of carbon that 
condenses into dust.  With the
solar metallicity for the Milky Way, we then obtain
\begin{equation}
{\rm dust}/{\rm HI}=1/(135\pm30)
\end{equation}
in agreement with the value adopted by models of dust to explain interstellar
reddening (Mathis et al.  1977; Weingartner \& Draine 2001) and also with
observations. Draine et al. (2007) estimated it to be 1/140 for the Milky
Way, and find for other galaxies that
this fraction varies between 1/100 and
1/400 with the median 1/190 for SINGS sample of galaxies.  
The fiducial value of 1/100 is usually taken as the dust/HI mass ratio
in the Milky Way.  
%This dust fraction explains empirical extinction in our Galaxy
%(Mathis et al.  1977; Weingartner \& Draine 2001).  
We assume that the heavy elements present in the interstellar 
matter condense into
dust with the fraction given by eq.  (\ref{eq:dustZ}), and
dust to HI ratio is universal in HI regions.

The HI survey (Zwaan et al. 2005; Zhang et al. 2008) 
and the H$_2$ survey (Keres et al. 2003)
give the cosmic value for the sum of 
atomic and molecular hydrogen to be
\begin{equation}
\Omega_{\rm HI+H_2}=5.35\pm0.87 \times 10^{-4} %10^{-3.27\pm0.06}, 
\end{equation}   
where ${\rm HI:H_2\approx 0.70:0.30}$.
With the assumption that dust resides in the HI region and the hydrogen
to dust ratio is universal
we estimate the global dust abundance in galactic discs, 
\begin{equation}
\Omega_{\rm dust~disc}=4.0\pm1.3 \times 10^{-6}  % 10^{-5.37}
\label{eq:discdust}
\end{equation} 
This is compared with the dust abundance estimated from
obscuration due to galaxies in a galaxy survey 
(Driver et al. 2007), $\Omega_{\rm dust~galaxy}=3\times 10^{-6}$. 

Dust may coagulate to form planets and may be depleted.  The amount we
estimated, however, is disturbed little by the planets formation,
while they are not entirely negligible.  Marcy (2005) estimated that
12\% of nearby FGK stars have detected Jupiter-like planets within 20
AU with the planet mass distribution $dN/dm\sim m^{-1}$.  In the
recent analysis Johnson et al. (2010) indicate that the fraction
depends on the mass of the central star and drops to 3\% for M dwarfs,
which dominate the stars in number.  Taking Figure 4 of Johnson et al.
we estimate that 0.052 planets formed per 1 $M_\odot$ of fuel
consumed. Here we extend the range of stars to span the full range of
M and A stars, by slightly extending the observed range which lies
between 0.25 to 2$M_\odot$.

The observed planets are of the Jupiter type, which is dominated by
hydrogen and helium gas. What concerns us here is dust used to form
the core of planets.  If we take the core accretion model for the
giant planet formation, the core mass is about $10M_\oplus$ per planet
(Mizuno 1980; Pollack et al. 1996; Rice \& Armitage 2003), in
agreement with the rocky core mass in the solar system planets, which
is $8, 10, 12M_\oplus$ for Jupiter, Saturn and Uranus (e.g., Lodders
\& Fegley 1998; Guillot 1999).  The mass density borne by planetary
cores that arose from coagulation of dust is $\Omega_{\rm
  planet}\simeq 8\times 10^{-9}$, only 1/500 the dust abundance in
eq.(\ref{eq:discdust}).  The knowledge concerning planets is still
immature and the estimates given here may be subject to revision in
the future.  Integrating the mass function of planets given above
around the Jupiter mass, we obtain a rough estimate of the mass
density of planets to be $\Omega_{\rm planet}\approx 7\times 10^{-7}$.

We now calculate the abundance of dust that is expected to be produced
from the stellar evolution.  With the Chabrier/Salpeter initial mass
function, the gas shed by stars is 0.60 times the mass locked into
stars with the resulting remnants given in section 1.  The majority of
stars have solar metallicity. We may consider that this is true even
at non-zero redshift, possibly except for some galaxies very early in
structure formation, as indicated by observations (e.g., Shapley et
al. 2004; de Mello et al. 2004) as well as demonstrated by a numerical
model for galaxy formation in the $\Lambda$CDM universe (Nagamine et
al. 2001), and the metallicity estimated from high redshift quasars.
The cosmic time is short at high redshift where these considerations
are more relevant. The dominant part of time relevant to star
formation is at low redshift $z<1-2$.  Motivated by these
observations, we may take the approximation that all stars have the
normal metallicity the same as at low redshift. 
We assume that the gas shed by stellar evolution has 
solar metallicity on average
whether gas taken into star formation is pristine or already somewhat
enriched. 
The amount of dust
produced by gas shed by stars is then
\begin{eqnarray}
\Omega_{\rm dust~produced}&=&\Omega_{\rm fuel}\times 0.38\times
\langle Z\rangle \eta\cr
&=&9.6\times 10^{-6}
\label{eq:SFdust}
\end{eqnarray} 
which is about 2.5 times larger than the global amount of dust in galactic
discs given in (\ref{eq:discdust}).
The total heavy element abundance calculated with this approximation correctly
reproduces the abundance observed at $z\approx0$, which in turn is
constrained by the total
energy output as explored by
extragalactic background light, as shown in FP04.

We also stress that a number of uncertainties in the calculation, such
as those in the initial mass function, mean metallicity, and the
dust/Z ratio, cancel when one compares eq. (\ref{eq:discdust}) with
eq. (\ref{eq:SFdust}), so that the error in the relative values is
not excessively larger.
The integral of
star formation rate still has a significant uncertainty, but it
is constrained by the observed amount of stars at $z=0$ and 
the fraction of stellar mass loss with the aid of the initial 
final mass relation, as
well as the energy output argument we quoted in section 1.  We may
take seriously
that the gap between the two values, 
expected amount of dust produced 
and that observed in the HI region of galaxies is  real.

\section{Dust in haloes and the closure of the dust entry}

This mismatch between the amount of dust that is ought to be produced
in stellar evolution and that is observed implies that either we miss
dust somewhere in the universe or dust produced is destroyed and
not all survives to now.  At the beginning, the latter may look natural
since lifetime of dust is thought to be not very long (Draine \&
Salpeter 1979; Draine 1995; Jones, Tielens \& Hollenbach 1996; Dwek
1998).  

We point out, however, that the recent detection of dust
reddening observed in the galaxy quasar correlation promotes us to
investigate the former case. Using a large quasar sample and yet an
even larger galaxy sample of the Sloan Digital Sky Survey, taking
advantage of its high precision multi-colour photometry, M\'enard et
al. (2010, hereafter MSFR) have found that the quasar light receives
reddening when it passes through the vicinity of galaxies. The colour
dependence of attenuation is in agreement with the dust extinction
curve known for the Milky Way, although the ratio of total to
selective extinction $R_V\approx 4.9\pm3.2$ is not well determined.
This implies that a large amount of dust is present around
galaxies ranging from 20kpc to a few Mpc, which are clearly beyond
galactic discs.

The projected surface density distribution of
dust follows $\sim r^{-0.8}-r^{-1}$ with $r$ the projected
distance from the centre of
the galaxy, similar to the galaxy mass distribution, to some 5Mpc.
The column density of dust in lines of sight 
is very small but when integrated over a large
volume this gives a substantial total
amount of dust: 
MSFR estimated the dust mass,
\begin{equation}
M_{\rm dust}\simeq5\times 10^7M_\odot
\label{eq:galdustmass}
\end{equation}
for $20{\rm kpc}<r<r_V$ with
$r_V$ the virial radius
for typical galaxies in the sample.

MSFR inferred $\Omega_{\rm halo~dust}\simeq
2.1\times 10^{-6}$, but this should be taken as a lower limit.
For the galaxy sample with median redshift of $\langle z\rangle=0.38$
the SDSS observation
samples galaxies down to $\approx0.25L^*$. The mean luminosity
of the sample is then  $\langle L\rangle\approx 0.7L^*$.
Assuming that
the amount of dust is proportional to luminosity,
the luminosity density  ${\cal L}_r\approx 2.2\times10^8hL_\odot$
(Mpc)$^{-3}$ means 
\begin{equation}
\Omega_{\rm halo~dust}\approx 2.6\times 10^{-6}
\label{eq:halodust}
\end{equation}
if (\ref{eq:galdustmass}) is used. 
MSFR shows that the distribution of dust
continues to be parallel to that of dark matter associated with
galaxy to several Mpc, on average beyond halfway to neighbouring galaxies.
 Remembering that the luminosity density, when multiplied
by M/L with M the bound mass, 
gives $\Omega\approx 0.15$ and it is compared with 
$\Omega\approx 0.27$ of global mean mass density, we conclude
that 45\% of mass is present outside the gravitationally-binding
radius of galaxies.  This is demonstrated using
the stacked surface matter density distribution derived from 
weak gravitational lensing signal, also in agreement with what we expect  
for the dark matter distribution in the CDM universe 
(Masaki, Fukugita \& Yoshida 2011, in preparation).
This suggestss that the amount of dust we estimated must be multiplied
by 1.8, including the distribution beyond the virial radius, 
to obtain the global abundance, if dust follows 
the dark matter distribution as
observationally indicated in MSFR.  This means that
$\Omega_{\rm galaxy+vicinity~dust}\approx 4.7\times 10^{-6}$,
if the vicinity of galaxies beyond the gravitationally bound
region is included.

The  uncertainty in the estimate of the amount of dust outside
the galaxies is admittedly large, but the order of the amount
we see  
indicates that the amount of dust in galactic discs must
be almost doubled for the total amount.  This means that the total amount
of dust 
\begin{equation}
\Omega_{\rm dust}\approx 9\times 10^{-6},
\label{eq:totaldust}
\end{equation}
which is close to what is expected from stellar evolution and the amount of
cooked fuels
(\ref{eq:SFdust}).

It has been discussed that dust is destroyed efficiently by sputtering
in the hot environment with the time scale,
$t\approx 10^5 {\rm yr} (n_{\rm H}/{\rm cm^{-3}})^{-1} (a/0.1 \mu{\rm
  m})$ for $T_{\rm eff}\sim 10^6$, the virial temperature of $L^*$
galaxies (Draine \& Salpeter 1979).  In the halo environment, $n_{\rm
  H}\sim 2\times 10^{-5}$ at 100kpc (Fukugita \& Peebles 2006), so that 
lifetime is longer than the age of the universe.  It seems likely that
dust in galactic haloes transported there from the galaxy during the
course of its formation and evolution 
may survive for the cosmological time.

Our argument indicates that the longevity of dust should also apply to
galactic discs. The total amount of dust we obtained is what is ought
to be produced.  We have no more fuels to yield extra dust. If dust
would be destroyed for some reasons, it must be replenished by
regeneration from interstellar matter.

We also remark that dust may be efficiently transported to outside
galaxies by galactic winds. For some galaxy samples it is observed
that a substantial fraction of the mass of formed stars is outflowed by
galactic winds (e.g., Heckman et al. 2002;  Pettini et al. 2002; 
Veilleux et al. 2005; Rupke
et al.  2005; Weiner et al. 2009), as also supported by plausible
arguments and simulations (e.g., Madau et al. 2001; Aguirre et al.
2001; Zu et al.). Whether a large part of dust is present in some
clumps, such as absorbing clouds (M\'enard et al, 2007), raises an
interesting problem, but in the present study, where we are concerned
with the coarse-grained mean abundance, we have no resolution to this
problem.

\section{Conclusions and discussion}

We have shown that the amount of dust expected in stellar evolution in
the cosmic time agrees with what we observe, when we take
account of dust that is uncovered in galactic haloes and in the
vicinity of galaxies from reddening of quasar light passing through
nearby galaxies (MSFR), which roughly doubles the total amount when
added to that observed in galactic discs.  Namely, we are observing
all dust produced in stellar evolution, which means that dominant
portion of dust should survive, at least effectively for cosmic time.
This contrasts to the conventional thought that lifetime of dust is
much shorter than the cosmic time.  This does not imply, however, that
dust be intact.  If it is destroyed in some processes in interstellar
or intergalactic space, it must be regenerated from the gas.  We
emphasise that we have no extra fuels to replenish destroyed dust,
meaning the closure of the cosmic energy inventory, which in turn
places a significant constraint on the physics of dust.  The dust
amount used to form cores of planets are only 1/500 in the disc.

If the dust production were ascribed solely to core collapse
supernovae, the total amount of dust we estimated means the productin
of $0.2M_\odot$ per core collapse.  This mass is consistent with the
prediction in some theoretical calculations (e.g., Kozasa et al. 2009;
Todini \& Ferrara 2001), but is 2 orders of magnitudes larger than is
actually observed at type II supernovae, $10^{-4}-10^{-3}M_\odot$. In
our consideration we do not necessarily mean that dust is produced at
the time or just after supernovae.  It may in part arise from mass
loss in the giant star phase before supernovae, or from gas ejected by
supernovae that may condense to dust later in the interstellar space
(Draine 2009).

The closure of the inventory tempts us to look at the partition more
closely as to the origin of dust. From the consideration of the final
versus initial stellar masses we expect that the stellar wind of stars
with $1-8M_\odot$ produces 56 \% of dust, where the dominant part
arises from the AGB phase (see Serenelli \& Fukugita 2007). Stars that
end with core collapse supernovae produce 43 \% of dust, including
both supernovae (and after supernovae) and mass loss in the
presupernova phase.  This means that the amount of dust produced in
core collapse is at most $0.08M_\odot$, smaller than the theoretical
estimate cited above.  Normalising the type Ia supernova rate to the
observed value at redshift $z\approx 0$ as in FP04, we infer that type
Ia supernova disrupt 5\% of white dwarfs which contribute 1\% of dust
at the time or some time after the explosion.  There is no evidence 
observed that type Ia supernovae yield dust.

Our consideration suggests that a substantial fraction of dust is
generated from the material lost in the burning phase of evolved
stars. The inventory gives useful circumstantial constraint on the production
and evolution of dust, although it does not directly suggest anything
concerning the mechanisms.

\vspace{6mm}
\noindent
{\bf ACKNOWLEDGMENT}
I would like to thank Bruce Draine and Peter Goldreich for encouraging
me to publish this work and for stimulating discussion and useful
comments improving the earlier version of this manuscript.  I also
thank Jim Peebles for a long-term collaboration on the cosmic energy
inventory, which motivated me to seek the \lq missing dust problem'.
I thank Brice M\'enard for discussion, for his work to find dust in
the vicinity of galaxies, and useful comments on the present manuscript.
I acknowledge the support of the Ambrose Monell Foundation (2010) and
the Friends of the Institute (2011) in Princeton and
Grant in Aid of the Ministry of Education in Tokyo.


\begin{thebibliography}{999}

\bibitem[Aguirre et al.(2001)]{2001ApJ...560..599A} Aguirre, A., Hernquist, 
L., Schaye, J., Weinberg, D.~H., Katz, N., 
\& Gardner, J.\ 2001, ApJ, 560, 599 

\bibitem[Asplund et al.(2005)]{2005ASPC..336...25A} Asplund, M., Grevesse, 
N., 
\& Sauval, A.~J.\ 2005, Cosmic Abundances as Records of Stellar Evolution and Nucleosynthesis, 336, 25 

\bibitem[Bahcall et al.(2001)]{2001ApJ...555..990B} Bahcall, J.~N., 
Pinsonneault, M.~H., \& Basu, S.\ 2001, ApJ, 555, 990 

\bibitem[Bernardi et al.(2010)]{2010MNRAS.404.2087B} Bernardi, M., Shankar, 
F., Hyde, J.~B., Mei, S., Marulli, F., 
\& Sheth, R.~K.\ 2010, MNRAS, 404, 2087 

\bibitem[Bouwens et al.(2010)]{2010arXiv1006.4360B} Bouwens, R.~J., et al.\ 
2010, arXiv:1006.4360 

\bibitem[Chabrier(2003)]{2003PASP..115..763C} Chabrier, G.\ 2003, PASP, 
115, 763 


\bibitem[de Mello et al.(2004)]{2004ApJ...608L..29D} de Mello, D.~F., 
Daddi, E., Renzini, A., Cimatti, A., di Serego Alighieri, S., Pozzetti, L., 
\& Zamorani, G.\ 2004, ApJL, 608, L29 

\bibitem[Draine(2009)]{2009ASPC..414..453D} Draine, B.~T.\ 2009, Cosmic 
Dust - Near and Far , 414, 453 


\bibitem[Draine(1995)]{1995Ap&SS.233..111D} Draine, B.~T.\ 1995, Ap\& SS, 233, 111 

\bibitem[Draine et al.(2007)]{2007ApJ...663..866D} Draine, B.~T., et al.\ 
2007, ApJ, 663, 866 

\bibitem[Draine 
\& Salpeter(1979)]{1979ApJ...231...77D} Draine, B.~T., \& Salpeter, E.~E.\ 1979, ApJ, 231, 77 

\bibitem[Driver et al.(2007)]{2007MNRAS.379.1022D} Driver, S.~P., Popescu, 
C.~C., Tuffs, R.~J., Liske, J., Graham, A.~W., Allen, P.~D., 
\& de Propris, R.\ 2007, MNRAS, 379, 1022 

\bibitem[Dwek(1998)]{1998ApJ...501..643D} Dwek, E.\ 1998, ApJ, 501, 643 

\bibitem[Fardal et al.(2007)]{2007MNRAS.379..985F} Fardal, M.~A., Katz, N., 
Weinberg, D.~H., \& Dav{\'e}, R.\ 2007, MNRAS, 379, 985 

\bibitem[Fukugita 
\& Peebles(2004)]{2004ApJ...616..643F} Fukugita, M., \& Peebles, P.~J.~E.\ 2004, ApJ, 616, 643 

\bibitem[Fukugita 
\& Peebles(2006)]{2006ApJ...639..590F} Fukugita, M., \& Peebles, P.~J.~E.\ 2006, ApJ, 639, 590 

\bibitem[Grevesse 
\& Sauval(1998)]{1998SSRv...85..161G} Grevesse, N., \& Sauval, A.~J.\ 1998, Space Science Reviews, 85, 161 

\bibitem[Guillot(1999)]{1999P&SS...47.1183G} Guillot, T.\ 1999, Planet. Sp. Sci., 47, 1183 

\bibitem[Heckman et al.(2000)]{2000ApJS..129..493H} Heckman, T.~M., 
Lehnert, M.~D., Strickland, D.~K., \& Armus, L.\ 2000, ApJS, 129, 493 

\bibitem[Heger et al.(2003)]{2003ApJ...591..288H} Heger, A., Fryer, C.~L., 
Woosley, S.~E., Langer, N., \& Hartmann, D.~H.\ 2003, ApJ, 591, 288 

\bibitem[Hopkins 
\& Beacom(2006)]{2006ApJ...651..142H} Hopkins, A.~M., \& Beacom, J.~F.\ 2006, 
ApJ, 651, 142 

\bibitem[Johnson et al.(2010)]{2010PASP..122..905J} Johnson, J.~A., Aller, 
K.~M., Howard, A.~W., \& Crepp, J.~R.\ 2010, PASP, 122, 905 


\bibitem[Jones et al.(1996)]{1996ApJ...469..740J} Jones, A.~P., Tielens, 
A.~G.~G.~M., \& Hollenbach, D.~J.\ 1996, ApJ, 469, 740 

\bibitem[Keres et al.(2003)]{2003ApJ...582..659K} Keres, D., Yun, M.~S., 
\& Young, J.~S.\ 2003, ApJ, 582, 659 


\bibitem[Kozasa et al.(2009)]{2009ASPC..414...43K} Kozasa, T., Nozawa, T., 
Tominaga, N., Umeda, H., Maeda, K., 
\& Nomoto, K.\ 2009, Cosmic Dust - Near and Far , 414, 43 


\bibitem[Lodders \& Fegley(1998)]{1998psc..book.....L} Lodders, K., \&
  Fegley, B.\ 1998, The planetary scientist's companion / Katharina
  Lodders, Bruce Fegley.~ New York : Oxford University Press,
  1998.~QB601 .L84 1998,

\bibitem[Madau et al.(2001)]{2001ApJ...555...92M} Madau, P., Ferrara, A., 
\& Rees, M.~J.\ 2001, ApJ, 555, 92 

\bibitem[Marcy et al.(2005)]{2005PThPS.158...24M} Marcy, G., Butler, R.~P., 
Fischer, D., Vogt, S., Wright, J.~T., Tinney, C.~G., 
\& Jones, H.~R.~A.\ 2005, Progress of Theoretical Physics Supplement, 158, 24 

\bibitem[Mathis et al.(1977)]{1977ApJ...217..425M} Mathis, J.~S., Rumpl, 
W., \& Nordsieck, K.~H.\ 1977, ApJ, 217, 425 

\bibitem[M{\'e}nard et al.(2010)]{2010MNRAS.405.1025M} M{\'e}nard, B., 
Scranton, R., Fukugita, M., \& Richards, G.\ 2010, MNRAS, 405, 1025 

\bibitem[M{\'e}nard et al.(2008)]{2008MNRAS.385.1053M} M{\'e}nard, B., 
Nestor, D., Turnshek, D., Quider, A., Richards, G., Chelouche, D., 
\& Rao, S.\ 2008, MNRAS, 385, 1053 

\bibitem[Mizuno(1980)]{1980PThPh..64..544M} Mizuno, H.\ 1980, Progress of 
Theoretical Physics, 64, 544 

\bibitem[Nagamine et al.(2001)]{2001ApJ...558..497N} Nagamine, K., 
Fukugita, M., Cen, R., \& Ostriker, J.~P.\ 2001, ApJ, 558, 497 

\bibitem[Nagamine et al.(2006)]{2006ApJ...653..881N} Nagamine, K., 
Ostriker, J.~P., Fukugita, M., \& Cen, R.\ 2006, ApJ, 653, 881 

\bibitem[Oohama et al.(2009)]{2009ApJ...705..245O} Oohama, N., Okamura, S., 
Fukugita, M., Yasuda, N., \& Nakamura, O.\ 2009, ApJ, 705, 245 

\bibitem[Ouchi et al.(2009)]{2009ApJ...706.1136O} Ouchi, M., et al.\ 2009, 
ApJ, 706, 1136 

\bibitem[Pettini et 
al.(2002)]{2002Ap&SS.281..461P} Pettini, M., Rix, S.~A., Steidel, C.~C., Hunt, M.~P., Shapley, A.~E., \& Adelberger, K.~L.\ 2002, Ap \& Space Sci, 281, 461 

\bibitem[Pollack et al.(1996)]{1996Icar..124...62P} Pollack, J.~B., 
Hubickyj, O., Bodenheimer, P., Lissauer, J.~J., Podolak, M., 
\& Greenzweig, Y.\ 1996, Icarus, 124, 62 

\bibitem[Rice 
\& Armitage(2003)]{2003ApJ...598L..55R} Rice, W.~K.~M., \& Armitage, P.~J.\ 2003, ApJL, 598, L55 

\bibitem[Rupke et al.(2005)]{2005ApJS..160..115R} Rupke, D.~S., Veilleux, 
S., \& Sanders, D.~B.\ 2005, ApJS, 160, 115 

\bibitem[Salaris et al.(2009)]{2009ApJ...692.1013S} Salaris, M., Serenelli, 
ApJ, Weiss, A., \& Miller Bertolami, M.\ 2009, , 692, 1013 

\bibitem[Serenelli 
\& Fukugita(2007)]{2007ApJS..172..649S} Serenelli, A.~M., \& Fukugita, M.\ 2007, ApJS, 172, 649 

\bibitem[Shankar et al.(2006)]{2006ApJ...643...14S} Shankar, F., Lapi, A., 
Salucci, P., De Zotti, G., \& Danese, L.\ 2006, ApJ, 643, 14 

\bibitem[Shapley et al.(2004)]{2004ApJ...612..108S} Shapley, A.~E., Erb, 
D.~K., Pettini, M., Steidel, C.~C., 
\& Adelberger, K.~L.\ 2004, ApJ, 612, 108 

\bibitem[Todini 
\& Ferrara(2001)]{2001MNRAS.325..726T} Todini, P., \& Ferrara, A.\ 2001, MNRAS, 325, 726 

\bibitem[Tremonti et al.(2004)]{2004ApJ...613..898T} Tremonti, C.~A., et 
al.\ 2004, ApJ, 613, 898 

\bibitem[Veilleux et 
al.(2005)]{2005ARA&A..43..769V} Veilleux, S., Cecil, G., \& Bland-Hawthorn, J.\ 2005, ARAA, 43, 769 


\bibitem[Weingartner 
\& Draine(2001)]{2001ApJ...548..296W} Weingartner, J.~C., \& Draine, B.~T.\ 2001, ApJ, 548, 296 

\bibitem[Weiner et al.(2009)]{2009ApJ...692..187W} Weiner, B.~J., et al.\ 
2009, ApJ, 692, 187 

\bibitem[Zhang et al.(2009)]{2009arXiv0902.2392Z} Zhang, W., Li, C., 
Kauffmann, G., Zou, H., Catinella, B., Shen, S., Guo, Q., 
\& Chang, R.\ 2009, arXiv:0902.2392 

\bibitem[Zu et al.(2010)]{2010MNRAS.tmp.1880Z} Zu, Y., Weinberg, D.~H., 
Dav{\'e}, R., Fardal, M., Katz, N., Kere{\v s}, D., 
\& Oppenheimer, B.~D.\ 2010, MNRAS, 1880 

\bibitem[Zwaan et al.(2005)]{2005MNRAS.359L..30Z} Zwaan, M.~A., Meyer, 
M.~J., Staveley-Smith, L., \& Webster, R.~L.\ 2005, MNRAS, 359, L30 


\end{thebibliography}
\end{document}